\documentclass{aa}  

\usepackage{graphicx}
\usepackage{epstopdf}
\usepackage{txfonts}
\usepackage{color}
\usepackage{booktabs}
\usepackage{multirow}
\usepackage{tabularx}
\usepackage{amsmath}
\usepackage{lipsum}
\usepackage{adjustbox}

\definecolor{linka}{rgb}{0.0,0.0,0.7}
\definecolor{linkb}{rgb}{0.0,0.0,0.5}
\definecolor{linkc}{rgb}{0.0,0.0,0.3}
\definecolor{cmt}{rgb}{0.5,0.0,0.0}
\definecolor{al}{rgb}{0.6,0.2,0.0}
\definecolor{mvn}{rgb}{0.4,0.4,0.0}
\definecolor{corr}{rgb}{0,0.0,0.4}
\definecolor{ref}{rgb}{0.4,0.0,0.0}
\definecolor{idea}{rgb}{0.4,0.0,0.0}
\usepackage[colorlinks=true,linkcolor=linkb,citecolor=linkc,filecolor=linka,urlcolor=linka,bookmarks=true,bookmarksopen=true,bookmarksopenlevel=3,plainpages=false,pdfpagelabels=true]{hyperref}
\usepackage{natbib}
\graphicspath{{./}{../figures/}{./figures/}}

\newcommand{\fig}[1]{Fig.~\ref{#1}} 
\newcommand{\Fig}[1]{Figure~\ref{#1}}

\newcommand{\sect}[1]{Sec.~\ref{#1}} 
\newcommand{\Sect}[1]{Section~\ref{#1}}

\newcommand{\sil}{Si\,\textsc{i}\,1082.7\,nm{} }
\newcommand{\he}{He\,\textsc{i}\,1083.0\,nm{} }

\newcommand{\colfigtwocol}[3][1.]{\begin{figure*}\centering
    \includegraphics[width=#1\linewidth,clip=TRUE]{#2}
    \caption{#3}
    \label{#2}
\end{figure*}}
\newcommand{\colfig}[3][1.]{\begin{figure}\centering
    \includegraphics[width=#1\linewidth,clip=TRUE]{#2}
    \caption{#3}
    \label{#2}
\end{figure}}

\begin{document} 

\title{Detection of emission in the \sil  line core in sunspot umbrae}
\titlerunning{Detection of emission in the \sil  line core in sunspot umbrae}

\authorrunning{D. Orozco Su\'arez et al.}

\author{D. Orozco Su\'arez\inst{1}
\and
C. Quintero Noda \inst{2}
\and
B. Ruiz Cobo\inst{3,4}
  \and
  M. Collados Vera\inst{3,4}
}

\institute{Instituto de Astrof\'isica de Andaluc\'ia (CSIC), Glorieta de la Astronom\'ia, 18008 Granada, Spain
\and
Institute of Space and Astronautical Science, Japan Aerospace Exploration Agency, Sagamihara, Kanagawa 252-5210, Japan
\and
Instituto de Astrof\'isica de Canarias, C/ V\'ia L\'actea, La Laguna, Spain
\and
Dept. Astrof\'isica, Universidad de La Laguna, E-38205, La Laguna, Tenerife, Spain
\\ \email{orozco@iaa.es}}
\offprints{D. Orozco Su\'arez, \email{orozco@iaa.es}}
\date{accepted: September 6, 2017}

\abstract
{Determining empirical atmospheric models for the solar chromosphere is difficult since it requires the observation and analysis of spectral lines that are affected by non-local thermodynamic equilibrium (NLTE) effects. This task is especially difficult in sunspot umbrae because of lower continuum intensity values in these regions with respect to the surrounding brighter granulation. Umbral data is therefore more strongly affected by the noise and by the so-called scattered light, among other effects.}
{The purpose of this study is to analyze spectropolarimetric sunspot umbra observations taken in the near-infrared  \sil line taking NLTE effects into account. Interestingly, we detected emission features at the line core of the \sil line in the sunspot umbra. Here we analyze the data in detail and offer a possible explanation for the \sil line emission.}
{Full Stokes measurements of a sunspot near disk center in the near-infrared spectral range were obtained with the GRIS instrument installed at the German GREGOR telescope. A point spread function (PSF) including the effects of the telescope, the Earth's atmospheric seeing, and the scattered light was constructed using prior Mercury observations with GRIS and the information provided by the adaptive optics system of the GREGOR telescope during the observations. The data were then deconvolved from the PSF using a principal component analysis deconvolution method and were analyzed via the NICOLE inversion code, which accounts for NLTE effects in the \sil line. The information of the vector magnetic field was included in the inversion process.}
{The \sil line seems to be in emission in the umbra of the observed sunspot after the effects of scattered light (stray light coming from wide angles) are removed. We show how the spectral line shape of umbral profiles changes dramatically with the amount of scattered light. Indeed, the continuum levels range, on average, from 44\% of the quiet Sun continuum intensity to about 20\%. Although very low, the inferred levels are in line with current model predictions and empirical umbral models. The \sil line is in emission after adding more that 30\% of scattered light so that it is very sensitive to a proper determination of the PSF. Additionally, we have thoroughly investigated whether the emission is a byproduct of the particular deconvolution technique but have not found any evidence to the contrary. Only the circular polarization signals seem to be more sensitive to the deconvolution strategy because of the larger amount of noise in the umbra. Interestingly, current umbral empirical models are not able to reproduce the emission in the deconvolved umbral Stokes profiles. The results of the NLTE inversions suggests that to obtain the emission in the \sil line, the temperature stratification should first have a hump located at about $\log\tau=-2$ and start rising at lower heights when moving into the transition region.}
{This is, to our knowledge, the first time the \sil line is seen in emission in sunspot umbrae. The results show that the temperature stratification of current umbral models may be more complex than expected with the transition region located at lower heights above sunspot umbrae. Our finding might provide insights into understanding why the sunspot umbra emission in the millimeter spectral range is less than that predicted by current empirical umbral models.}

\maketitle

\section{Introduction}

In the literature, there are several proposed empirical models specific to sunspot umbrae \citep{1981phss.conf..235A,1986ApJ...306..284M,1988A&A...189..232O,1994A&A...291..622C,1994ASIC..433..169S,2007ApJS..169..439S,2009ApJ...707..482F}. Most of these models have been obtained empirically with observational data taken at several wavelengths (UV, visible, etc). However, although modeling the temperature and vector magnetic magnetic field stratification in the photospheric part of sunspot umbrae is relatively easy, this is not the case for the chromosphere and layers above. In these regions, most observed spectral lines suffer from non-local thermodynamic equilibrium effects (NTLE), which makes it difficult to determine the height variation of several thermodynamic parameters accurately. For this reason, while the different available models look similar to each other in the photosphere these models significantly differ at higher layers, especially at the chromosphere (see \fig{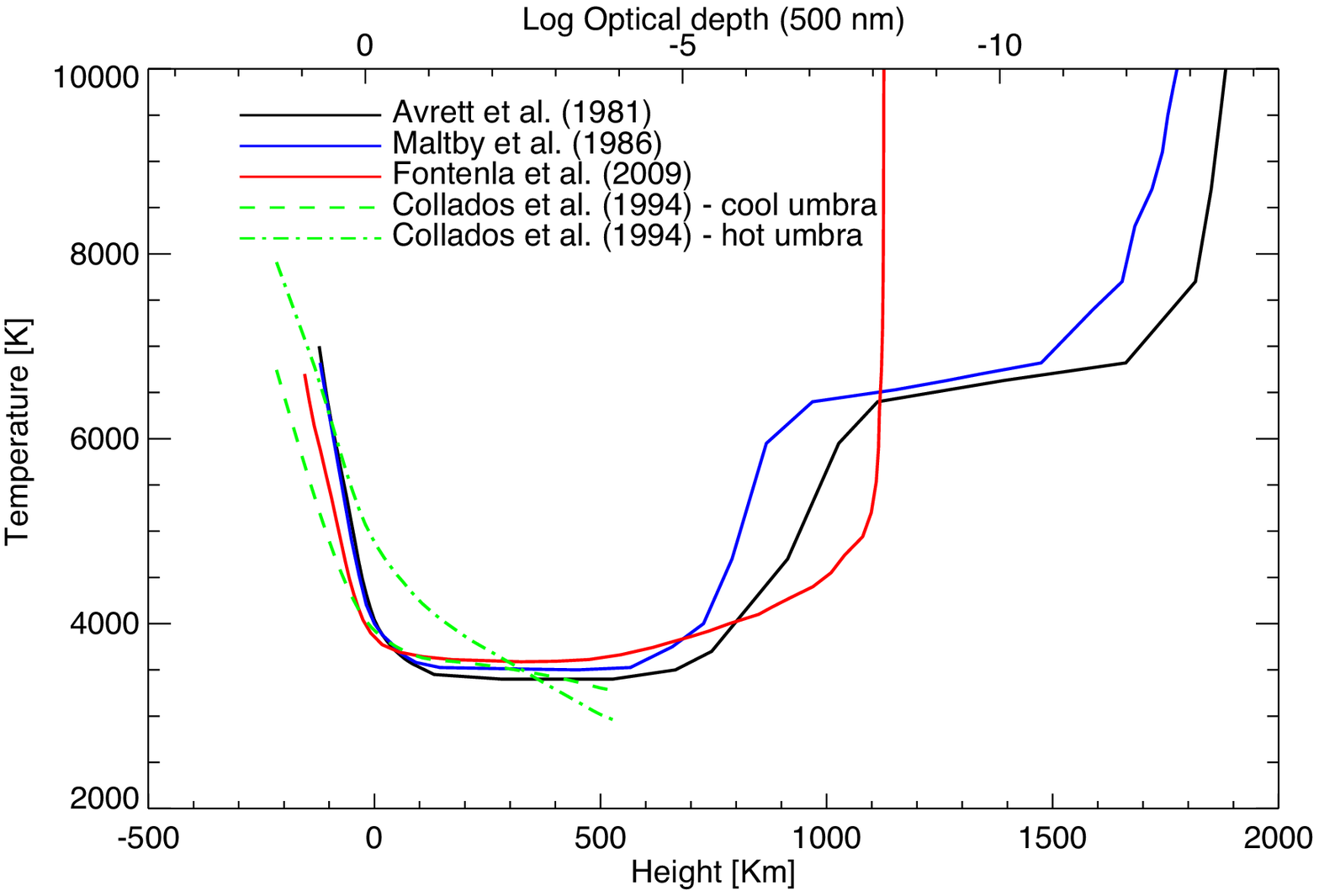}). To properly model the thermodynamic parameters at these layers, spectral lines sampling of those heights should be included in the modeling process. This particular problem has been already addressed \citep[see, e.g., review of ][]{2003A&ARv..11..153S}. The most recent model we find in the literature is that from \cite{2007ApJS..169..439S}. These authors inferred a sunspot umbral model from the analysis of several chromospheric Ca\,\textsc{ii}\, and photospheric Fe\,\textsc{i}\, lines. They used an inversion code that is able to synthesize spectral lines in NLTE conditions. Other umbral models do not take into account the chromosphere \citep{1994A&A...291..622C}.

Very recently, \cite{2016ApJ...816...91I} \citep[see also][]{2015ApJ...804...48I} have found,
using the Nobeyama Radioheliograph \citep[NoRH;][]{1994IEEEP..82..705N}, that the observed umbral brightness temperature is the same as that of the quiet Sun at 34 Ghz (corresponding to an 8.8 mm range). Their findings are inconsistent with current model predictions, which suggest that sunspot umbrae should be brighter than the quiet Sun at the millimeter spectral range. Indeed, none of the umbral models available to the solar community are able to explain these radio observations \citep{2016ApJ...816...91I}. From the results of Iwai et al. it seems that the temperature should start increasing at a lower geometrical height, so that the minimum temperature region should be located lower in the atmosphere when above sunspot umbrae. As we stated above, constructing sunspot umbrae atmospheric models for the upper photosphere and low chromosphere is not straightforward since it entails the inclusion of NLTE spectral lines in the modeling process. So far, only \cite{2007ApJS..169..439S} have included chromospheric lines. In this paper, we present a sunspot observation taken in the \sil line ($4s^3P^o_2 - 4p^3P_2$, with Land\'e factor 1.5, excitation potential 4.95 eV and $\log\,gf = 0.363$\footnote{Taken from \cite{Borrero2003}}) at high spatial resolution ($\approx$~0\farcs3) with the GRegor Infrared Spectrograph \citep[GRIS;][]{2012AN....333..872C}. Remarkably, the core of the \sil line is mainly formed in NLTE conditions \citep{Bard2008}. The novelty of these observations is that, once the stray-light  effects, including the telescope, atmospheric seeing, and scattered light, are removed from the observations, the \sil line appears in emission at the umbral core. The analysis of these observations, as we detail throughout this paper, suggests that the minimum temperature region should be located at lower heights, at least for this sunspot, with a temperature hump located at about $\log\tau=-2$. Should this result be extended to average sunspot umbrae models, it could provide a natural explanation of the controversy between the present empirical models and the radio observations at the millimeter range. Moreover, it could also settle a new paradigm regarding umbral stratification. In \sect{grisobs} we describe the observations and explain the deconvolution process. In \sect{emission} we detail the \sil emission profiles and discuss the results of the deconvolution process, and in  \sect{NLTE} we cover the inversion of the data under NLTE conditions. Finally, in \sect{discussion} we summarize the results and highlight the main conclusions and possible impacts. 

\colfig{fig1.eps}{Variation of the temperature with atmospheric height corresponding to various atmospheric models available in the literature. In particular, the E model of \cite{1981phss.conf..235A}, the models by \cite{1986ApJ...306..284M} and \cite{2009ApJ...707..482F} that cover from the photosphere to the chromosphere, and the model of \cite{1994A&A...291..622C} (cool umbra and hot umbra) that only covers the photosphere. The optical depth scale has been calculated using the \cite{1986ApJ...306..284M} atmospheric model.}

\section{Sunspot observations and PSF correction\label{grisobs}}

\colfigtwocol{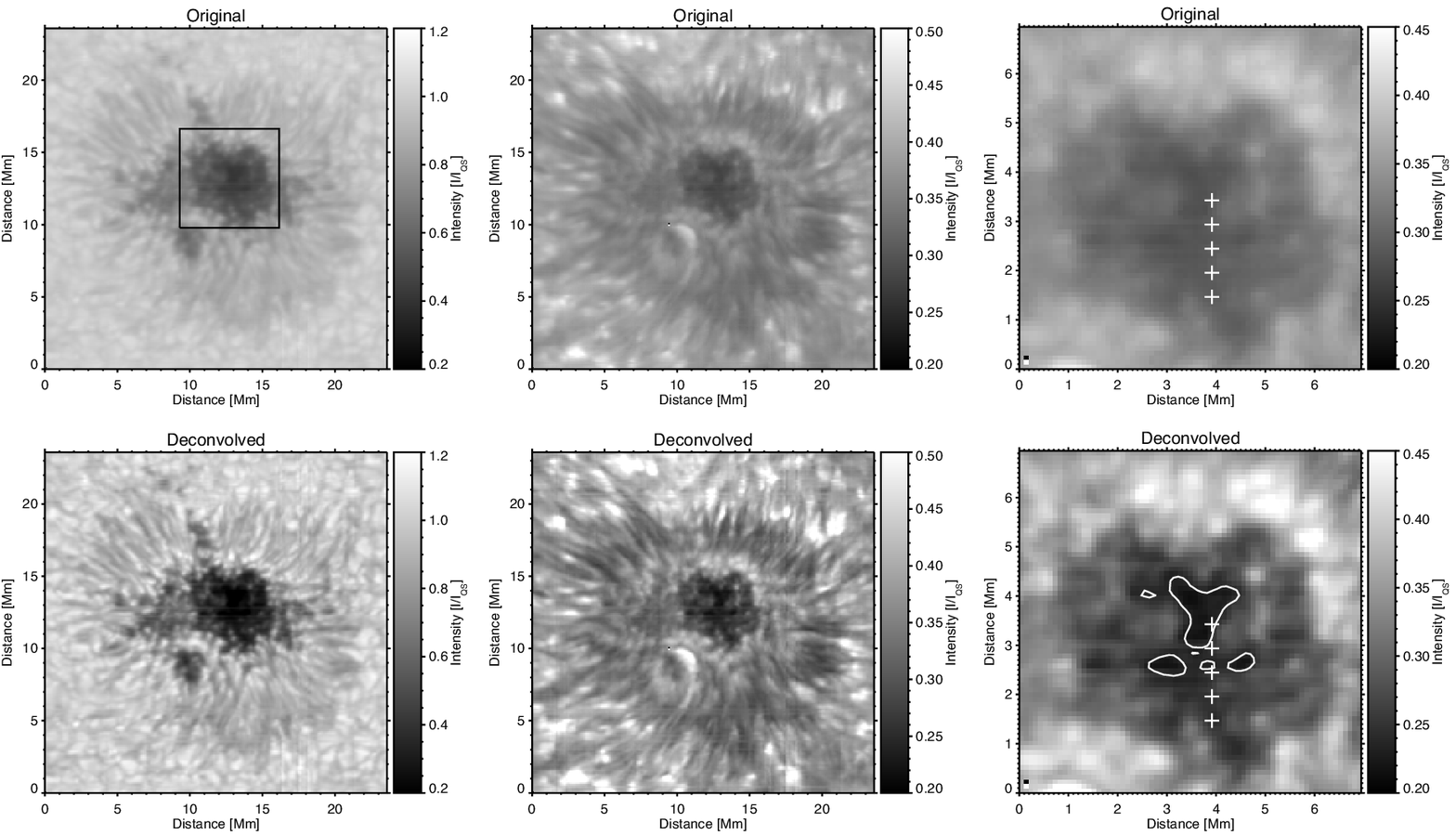}{Intensity maps of the observed NOAA 12096 sunspot corresponding to the continuum (left panels) and to the \sil line core (center panels). The top panels represent the original data and the bottom panels the reconstructed data. The right-most panels show a zoom over the original (top) and reconstructed (bottom) sunspot umbra in the continuum (see the square in the top left panel). The contours outline the region where the \sil continuum is below 25\% of the average quiet Sun.}\label{three}

The sunspot observations were taken on 27 June 2014 with the GRIS spectrograph of the $1.5$ meter German GREGOR telescope \citep{2012AN....333..796S} at the Observatorio del Teide (Tenerife, Spain). The sunspot (NOAA 12096) was located close to the disk center ($x=-72,5$\arcsec{}, $y=148$\arcsec{}). The GRIS spectograph is equipped with an infrared camera \citep[TIP II;][]{2007ASPC..368..611C} and a polarization modulation package. Thus, the four Stokes parameters were measured with TIP-II around the near-infrared spectral region containing the \he multiplet and the  \sil photospheric line and an atmospheric water vapour line at 1083.21~nm. The size of the scanned region is 32\arcsec{}x63\farcs3. The four Stokes parameters were recorded by moving the slit (0\farcs135 wide) in 240 step positions with a step size of 0\farcs135. The effective pixel size of the data is 0\farcs135. The total effective exposure time per slit position was about 1.5\,seconds, so that the full scan lasted 6 minutes from 10:06~UT until 10:12~UT.

The standard data reduction software was applied to the GRIS data, which include dark current and flat-field correction, polarimetric calibration, and cross-talk removal (see Collados et al.; in preparation). The wavelength calibration was carried out by fitting the observed spectral region to the McMath-Pierce Fourier Transform Spectrometer (FTS) spectrum \citep{1991aass.book.....L,1993aps..book.....W}, taken the \sil line of the average quiet Sun and the water vapour line at 1083.21~nm of the atlas as reference wavelengths. The resulting spectral sampling is 54.3\,m\AA{}$/$pixel, after averaging three pixels in the wavelength dimension. This operation also allowed us to correct the continuum level of the data; that is, we removed the low frequency oscillation of the continuum. The spectral resolution of GRIS at 1083~nm should be around 120~m\AA\/ accordingly to \cite{2016A&A...596A...8J}. More information about these procedures can be found in \cite{2016A&A...596A...4F} and \cite{2016A&A...596A...2B}.

The seeing conditions during the observations were good. The GREGOR Adaptative Optics System \citep[GAOS;][]{2012AN....333..863B} was running continuously. The alto-azimuthal mounting of the telescope introduced an image rotation\footnote{Presently, GREGOR is equipped with an image derotator.} smaller than $0.5 [^o/min]$ at 10 UT \citep{2016A&A...596A...4F}, so that this effect is almost negligible in our data set. Finally, we removed an 8\% of spectrograph scattered light from the observations with the procedure described in \cite{2016A&A...596A...4F} and \cite{2016A&A...596A...2B}. 

The observations are affected by the telescope optics, the Earth's atmosphere (seeing), and  scattered light. This stray-light $\mathrm{PSF}(r)$ can be modeled by the sum of two Gaussian components \citep{2016A&A...596A..59F} as follows:
\begin{equation}
\mathrm{PSF}(r) = \alpha G_1(r,\sigma_1) + (1-\alpha) G_2(r,\sigma_2)
,\end{equation}
where $G_1(r,\sigma_1)$ and $G_2(r,\sigma_2)$ are two Gaussian functions representing the spatial resolution of the telescope and atmospheric seeing and the light scattered over large angles. The values  $\sigma_1$ and  $\sigma_2$ stand for the rms width of both contributions while $\alpha$ is the relative contribution of the Gaussians, and $r$ represents the radial distance $r^2 = x^2 + y^2$. Both contributions were determined experimentally with observations of the Mercury transit in front of the Sun on 2016 May 9 and GREGOR adaptive optics. Weather conditions were bad during the Mercury transit but the acquired data were good enough to allow the determination of the instrumental stray-light contribution to the GREGOR PSF, which is  $\sigma_2 = 5.030\pm0.001$\arcsec{} and $\beta = (1 - \alpha) = 42\%$. Once the contribution of the second Gaussian is fixed, we used the information provided by the GAOS system to compute the contribution of the telescope and the Earth's atmosphere to the GREGOR PSF during our observations. In our case, the estimated Fried parameter at 500~nm was $r_0 = 15.4$~cm with an rms variation of  $\pm0.6$~cm, which corresponds to a $\sigma_1 = 0.24$\arcsec{}, in the near-infrared at 1083~nm. In reality, the PSF varies continuously during the 6 minutes observation period, mostly due to the changing conditions of the Earth's atmosphere (turbulence). For the calculations above we took the average Fried parameter over the 6 minute period. The rms variation of the fried parameter was small though, as it was approximately $\pm0.6$~cm during the observations. 

We used the principal component analysis (PCA) deconvolution method described by \cite{2013A&A...549L...4R} and \cite{2015A&A...579A...3Q} to remove the stray light. This deconvolution technique has already been applied to GRIS data by \cite{2016A&A...596A...2B}, who describe how the method works in detail. In our case, we limited the information to the first 12 PCA coefficients and performed only 20 iterations within the deconvolution process.
We applied the PCA deconvolution method to the spectral region containing only the \sil line. It is important to note that the PCA deconvolution method is applied to ``solar images'' constructed from  the slit images (240 in our case). In this sense, the solar scene and the PSF vary along the scan direction, where this issue is one of the main drawbacks of image deconvolution techniques applied to spectroscopic data. However, we find that the most important effect comes from the contribution of the second Gaussian of the PSF (stray light), which we assume to be constant during the observations. The amount of stray-light contribution coming from the second Gaussian depends on the solar scene as well, but we can assume that the evolution of the solar scene along the scanning direction has negligible impact on the stray-light contribution because this contribution comes from wide angles. Finally, the deconvolution process is applied independently to the four Stokes parameters. 

\Fig{fig2.eps} shows continuum Stokes I/I$_\mathrm{QS}$ and \sil line core maps of the observed sunspot before and after applying the PCA deconvolution, where I$_\mathrm{QS}$ is the average quiet Sun intensity at the continuum. It can be seen how the contrast and image sharpness has increased considerably after the deconvolution of the data, especially at the line core. The effect in the sunspot umbra is also evident; for instance, the intensity at the umbra has clearly diminished after the deconvolution. However, as we see in detail in the next section, the most conspicuous effect can be found in the spectra recorded at the sunspot umbra. In the original data, the \sil line was always in absorption, while it becomes in emission after the deconvolution. The contours in \fig{fig2.eps} outline the area where the sunspot umbral intensity is lower than 25\%\/ (compared to the quiet Sun). Within that area, the core of the \sil line appears in emission.

\section{\sil profiles in emission\label{emission}}

\colfigtwocol{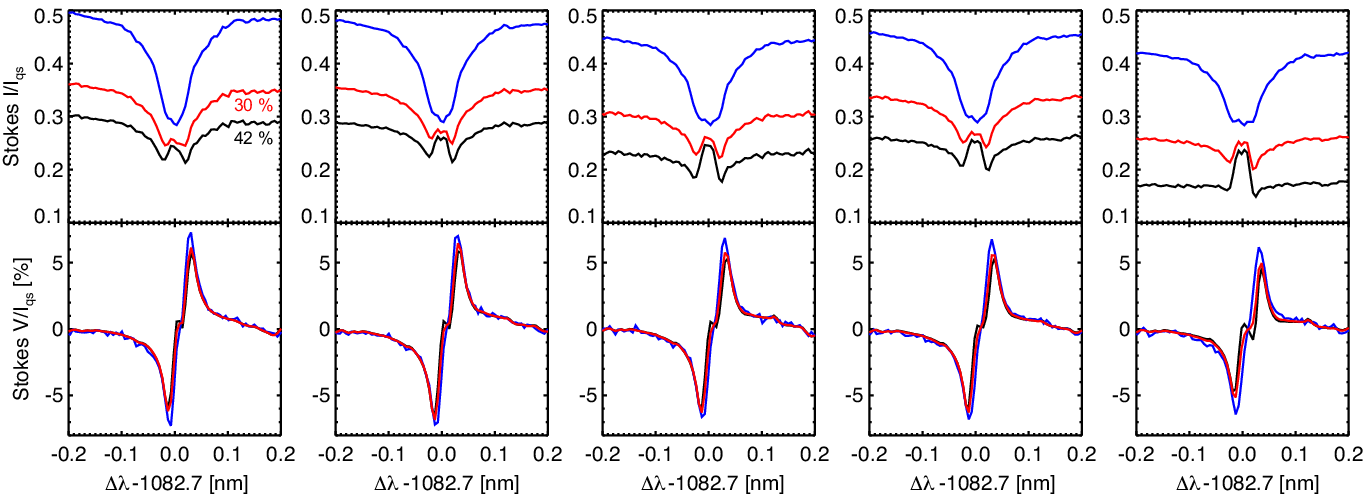}{Variation of the original (blue) and reconstructed Stokes I/I$_\mathrm{QS}$ and V/I$_\mathrm{QS}$ profiles as we move from the inner penumbral boundary (first column) to the center of the umbra (last column). The Stokes I profiles in black correspond to the deconvolved profiles taking into account 42\% of stray light while for the red profiles we only accounted for 30\% of the stray light. For simplicity, we only represent five pixels. The location of the pixels can be found in the right panels of \Fig{fig2.eps}. The first profile corresponds to the lowest point.}

\Fig{fig3.eps} shows how the shape of Stokes I/I$_\mathrm{QS}$ and V/I$_\mathrm{QS}$ changes from the inner penumbral boundary to the umbral core for the estimated 42\% of stray light and for 30\% of stray light for comparison. In the original data, Stokes I/I$_\mathrm{QS}$ never goes below 40\%\/ of the intensity (relative to the quiet Sun average intensity). Moreover, the profiles are rather regular and show three tiny peaks at the line core, basically as a result of the Zeeman splitting under the presence of a strong magnetic field; the left and right absorption peaks correspond to the $\sigma$ components while the central absorption peak correspond to the $\pi$ component. The stray-light contamination contributes to the central absorption peak, since the stray light has, in general, the shape of Stokes I in the absence of a magnetic field. The $\pi$ component is absent if the vector magnetic field is strictly pointing toward or away from the observer. Similarly, Stokes V/I$_\mathrm{QS}$ shows the prototypical anti-symmetrical shape with two lobes of opposite sign. However, the deconvolved profiles show a completely different scenario. In Stokes I, the intensity has dropped considerably, reaching values below 20\% at the umbral core where the profile clearly shows the $\sigma$ components. The line core intensity is much greater than the intensity at the wavelength location of the $\sigma$ components. As we see later, we ascribe such intensity excess to line emission and refer to it as emission peak. In the profiles located at the umbral core (rightmost profile in \fig{fig3.eps}), the emission peak is clearly discernible, flanked by two tiny absorptions. Regarding Stokes V/I$_\mathrm{QS}$, their amplitudes with respect to the quiet Sun decrease only slightly after the deconvolution. Interestingly, the Stokes V/I$_\mathrm{QS}$ profiles always show a reversal in the zero-crossing wavelength in all cases although it is much more prominent at the umbral core. The intensity reversal is not recovered when the straylight is only reduced to 30\%. Regarding Stokes Q/I$_\mathrm{QS}$ and Stokes U/I$_\mathrm{QS}$ (not shown), their amplitude is very small, i.e., below 1\% for the former and negligible (within the noise level) for the latter. This discards the magneto-optical effects as an explanation for the Stokes V/I$_\mathrm{QS}$ zero-crossing reversal.

Overall, the effects of the deconvolution process in the Stokes profiles (especially for Stokes I/I$_\mathrm{QS}$) are rather dramatic. For this reason, we evaluated whether the deconvolution worked properly by looking at possible side effects ascribed to the PCA deconvolution strategy or to an inaccurate determination of the PSF of the telescope. The first thing to check is the recovered continuum intensity levels, which have dropped considerably with respect to the original values. In the darkest part of the umbra,  the continuum intensity changes from the 42\% in the non-deconvolved data to about 17\% after the reconstruction as suggested by the Stokes I/I$_\mathrm{QS}$ shown in the rightmost panel of \fig{fig3.eps}. However, on average, the intensity variation in the darkest area of the umbra (outlined in \fig{fig2.eps}, right panel) goes from 44\% of the continuum in the non-reconstructed data to about 20\%. If we take into account the whole umbra, which is defined as the region in the reconstructed data that has a continuum intensity below 40\%, the continuum intensity decreases from 50\% to about 30\% on average. Current models for a sunspot umbra  of about 2.5~Mm radius\footnote{In any case, the exact dependence is not only with umbral radius but also with the magnetic flux, the location on the solar disk, and the umbral structure, among others.} suggest about 25\% of continuum intensity (following \cite{refId0}, see their Figure 5) for visible wavelengths. The continuum intensity also varies with wavelength, which is typically larger for longer wavelengths. For instance, \cite{1986ApJ...306..284M} suggest continuum values of about 22\% for 1~$\mu m$ (see their figure 2). The continuum intensities recovered in the GREGOR data are slightly lower than the predictions for the darkest part of the umbra. To shed light into this issue we show, in  \fig{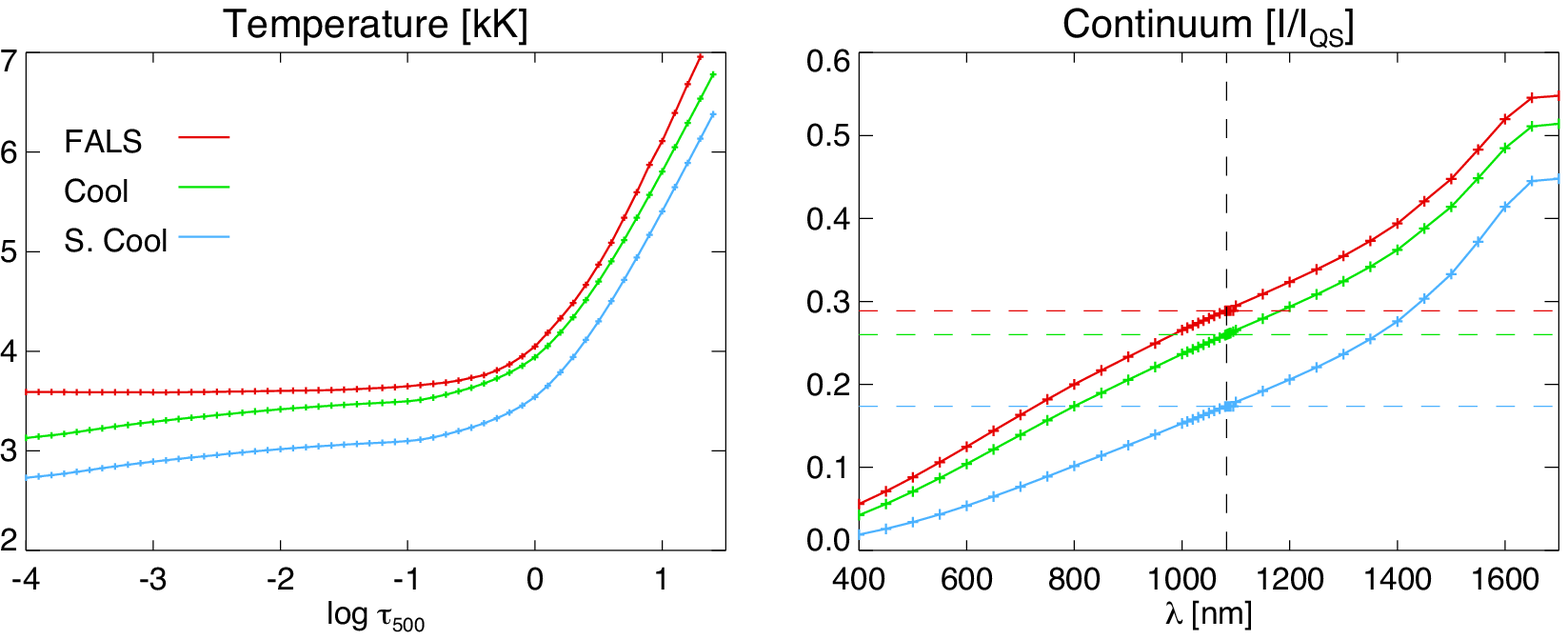}, three different atmospheric models, i.e., the FALS, the cool model of \cite{1994A&A...291..622C}, and a \emph{super-cool} model, along with the variation of the corresponding intensity values at the continuum with wavelength. The \emph{super-cool} model was tailored from the cool model to get continuum values of about 17\% in the 1083~nm region. To get such continuum intensity values in the infrared, we needed to shift the temperature down about 400~K. Thus, the model would be rather cool. Interestingly, the intensity values that would be recovered with this model for visible wavelengths (around 630~nm) would be about 7\%, which is in line with the intensity values reported in Hinode observations with deconvolved data \citep{2013A&A...549L...4R}. Remarkably, the cool model of   \cite{1994A&A...291..622C} already predicts intensity values of about 30\%, which is in line with the (averaged) intensity values for the deconvolved sunspot umbra of around 30\%.

\colfigtwocol{fig3b.eps}{Variation of umbra continuum intensity with wavelength (right panel) for three different models (left): the FALS, the cool model of \cite{1994A&A...291..622C}, and a \emph{super-cool} model obtained from the cool model. The vertical line shows 1083 nm wavelength while horizontal lines pinpoint the corresponding continuum intensity values at 1083~nm for the three models.}

Therefore, we believe that a $\beta = 42\%$ of scattered light for the GREGOR telescope seams reasonable. Remarkably, if we reduced the stray-light contribution to 30\%, we would obtain continuum intensity values significantly higher than those predicted in the models, for the darkest part of the umbra (about 30\% on average). In \Fig{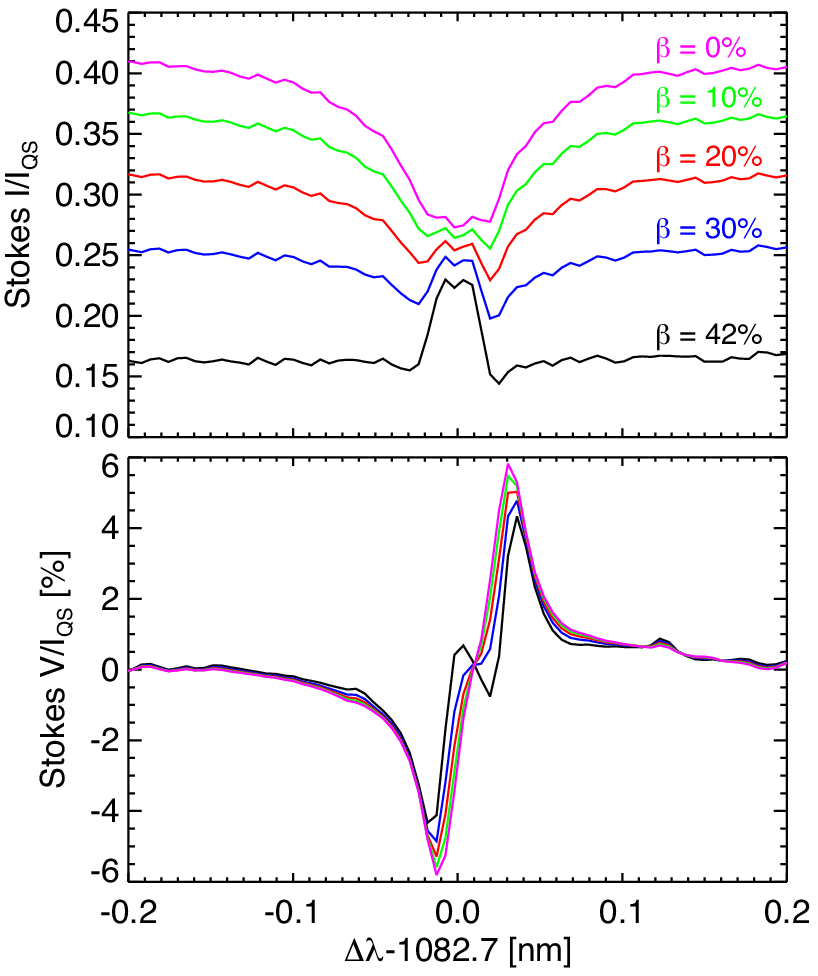} we represent the variation of the Stokes I/I$_\mathrm{QS}$ and V/I$_\mathrm{QS}$ shapes with increasing amount of scattered light ($\beta = 0\%, 10\%, 20\%, 30\%, \mathrm{and}\,\, 42\%$). For $\beta=30\%$ and $42\%$, the recovered Stokes I/I$_\mathrm{QS}$ can be easily associated with line emission. We only recovered a Stokes V/I$_\mathrm{QS}$ line reversal at the zero-crossing point with values above
$\beta=30\%$. For lower $\beta$ values, it is not possible (by eye) to guess whether  the line is in emission or if it is just the Zeeman splitting of the line. The figure also demonstrates the fact that the contribution of the second Gaussian (scattered light) to the global PSF is the most critical one. Changing the width of the first Gaussian (core of the PSF) basically affects to the spatial sharpness only. 

Furthermore, when we recovered the emission feature in Stokes I/I$_\mathrm{QS}$, we simultaneously recover a line reversal around the zero-crossing wavelength in Stokes V/I$_\mathrm{QS}$. Since the PCA deconvolution method is applied separately to the Stokes I/I$_\mathrm{QS}$ and Stokes V/I$_\mathrm{QS}$ profiles, it could be argued that the emission feature is not due to errors within the deconvolution process itself. Thus, it makes sense to check whether the recovered Stokes I/I$_\mathrm{QS}$ and V/I$_\mathrm{QS}$ are compatible. We discuss this issue in \sect{nlte_synt}.

Another source of error may be found in the fact that the PCA deconvolution method uses a database to reconstruct the Stokes profiles. Such a database is constructed using the very same (observed) Stokes profiles. Thus, in the case of Stokes I, the database does not contain Stokes I/I$_\mathrm{QS}$ profiles in emission. But, as one may erroneously think, this does not limit the PCA method to deliver profiles with emission signatures since they are constructed from linear combinations of all available profiles in the database, i.e., 12 in our case. The exact same situation takes place for Stokes V/I$_\mathrm{QS}$, although Stokes V/I$_\mathrm{QS}$ profiles with opposite polarity are available in the database. In any case, and to discard any other effects within to PCA deconvolution process (such as that previously mentioned), we repeated the data deconvolution using a completely different method known as the maximum entropy method \citep{1992ApJ...386..293H}. This method is based on fast Fourier transforms, so it is rather different from PCA. The results from the deconvolution, shown in \Fig{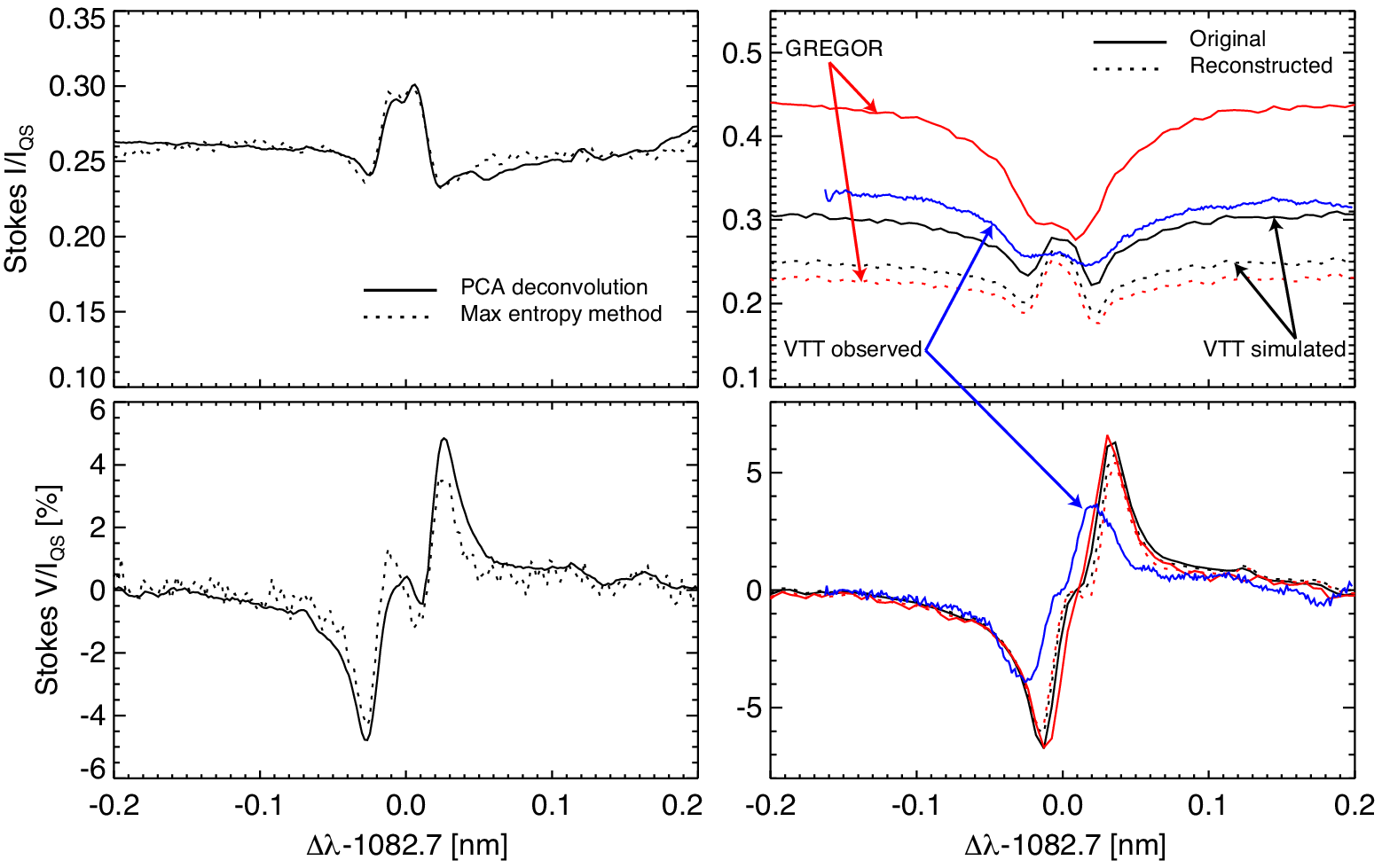}, are exactly the same for the sunspot umbra. The only noticeable difference is that the deconvolved profiles using the maximum entropy method are much more affected by the noise inherent to the deconvolution process, and actually an advantage of using PCA is that we can keep the noise at low levels since PCA also acts as a noise filter. 

We also varied several parameters within the PCA deconvolution method concerning the number of iterations and number of profiles used to construct the database but none of these parameters affected the deconvolution results in a significant way, i.e., we still get the line core emission (see Appendix A). Interestingly, we see small variations, although at the very limit around the Stokes V/I$_\mathrm{QS}$ zero-crossing point. This issue is discussed in detail in \sect{nlte_synt}.

Finally, to explore the deconvolution method further, we applied it to the red side of the observed spectral region containing the Ca\,\textsc{i}\,1083.34\,nm \citep[see, e.g.,][]{2016A&A...596A..59F} and the Na\,\textsc{i}\,1083.48\,nm lines both located at the sides of a Ti\,\textsc{i}\,1083.36\,nm atmospheric line and the Ca\,\textsc{i}\,1083.9\,nm \citep[see, e.g.,][]{2016A&A...596A...8J,2017A&A...599A..35J}. The results of the deconvolution (see Appendix B for a detailed description) suggest that these lines are not in emission in the Sunspot umbra. This may be simply because these lines are weak and do not reach the upper photosphere or low chromosphere.

\colfig{fig4.eps}{Variation of Stokes I/I$_\mathrm{QS}$ and V/I$_\mathrm{QS}$ profiles with the amount of scattered light $\beta$ from 0\%\/ to 42\%. The profile corresponds to that shown in the fifth panel on \Fig{fig3.eps}.}\label{evolucion}

\colfig{fig5.eps}{Left panels represent an umbra Stokes I/I$_\mathrm{QS}$ and V/I$_\mathrm{QS}$ profiles reconstructed using the PCA deconvolution and maximum entropy methods. The right panels show an umbra Stokes I/I$_\mathrm{QS}$ and V/I$_\mathrm{QS}$ profile as seen by GREGOR and the German VTT. The solid line represents the non-reconstructed data while the dotted line stands for the reconstructed data. The black line corresponds to the German VTT (simulated), red to the GREGOR telescope, and blue to the observed VTT profile (see text).}

The question remains as to why these emission profiles have not been seen before,  for instance, at the German Vacuum Tower Telescope (VTT) of the Observatorio del Teide. The TIP-II instrument has been operating at the VTT for many years and there are many observations of sunspots with that telescope and instrument. The main differences between the VTT and the GREGOR telescopes are the size of the primary mirror (70~cm versus 1.5~m) and that the VTT has, in principle, less scattered light, i.e., about 10\%\/ \citep{2011A&A...535A.129B}. We carried out the following test to see how the observed and reconstructed GREGOR umbral profiles would have been seen with the German VTT. We took the reconstructed sunspot data (as observed in GREGOR) and degraded these data with a PSF and a pixel sampling corresponding to the German VTT (assuming same seeing conditions and 10\%\/ scattered light). The results are shown in \Fig{fig5.eps} (right panels). It can be seen that the shape of the simulated Stokes I profile (solid black) resembling the German VTT is very similar to the observed and reconstructed GREGOR profile (dotted red). Moreover, the line core intensity almost reaches that of the continuum level. The large difference between the non-reconstructed GREGOR profile (solid red) and the corresponding simulated VTT profile (solid black) can be directly ascribed to the different amount of scattered light (42\% in GREGOR versus 10\% in the VTT) and, to a lesser extent, to the spatial resolution. When we compared the deconvolved GREGOR profile with the simulated VTT profile (red and black dotted curves, respectively) the main difference between them is simply due to the smaller spatial sampling (and resolution) of the German VTT (assuming same atmospheric seeing conditions). This simplistic simulation suggests that umbral emission profiles should have been seen earlier at the VTT, without applying any deconvolution method to the data. For this reason, we reviewed previous VTT data, looking for a sunspot at disk center with similar size and good observing conditions. In particular, we made use of a data set taken on 27 September 2011 with a 0\farcs32 spatial sampling. The radius of the spot was about 4.8~Mm, which is a bit larger than that presented here and was located at $x=228$\arcsec{}, $y=90$\arcsec{}. A profile taken from the umbral core of the sunspot contained in this data set can be seen in \Fig{fig5.eps} (solid blue). Remarkably, the continuum level\footnote{Typical continuum levels in sunspot umbra recorded in the VTT are around 0.4-0.5.} of the observed Stokes I/I$_\mathrm{QS}$ at the VTT is very close to the simulated Stokes I/I$_\mathrm{QS}$ with the GREGOR observations, i.e., about 0.3, so a wide-angle scattered light of 10\% at the VTT seems to be a good lower limit. The main difference between the simulated and observed intensity profile is that the latter shows less contrast in the wavelength dimension, i.e., it is shallower. This difference could be perfectly ascribed to the spectral resolution and a non-negligible spectral scattered light, which has not been accounted for in the observed VTT profile. Regarding Stokes V/I$_\mathrm{QS}$, although it has a smaller amplitude, its shape is very similar to the simulated Stokes V/I$_\mathrm{QS}$ around the zero-crossing wavelengths (solid black). We believe that if we apply a deconvolution on these data, we would recover an emission profile as we do with the GREGOR data. We can argue that, at the limited spatial resolution of the German VTT, the amount of scattered light is critical to the point where if the scattered light is larger than 10\%, the emission signals would be significantly suppressed. To summarize, we think that the German VTT telescope was at the limit of the detection of the emission signals in the \sil line. 

\section{NLTE treatment of the \sil line \label{NLTE}}

In this section we try to infer the physical conditions on the umbra atmosphere of the sunspot that gives rise to the \sil line core emission. To this end, we generate synthetic \sil Stokes profiles and compare these with the observed profiles, putting special emphasis on the line core emission. This forward modeling  process, known as ``inversion'' of Stokes profiles, allows us to infer the temperature and magnetic field vector stratification in this particular sunspot umbra. The \sil line is mainly formed in the photosphere but the line is very deep and its core is dominated by NLTE \citep[e.g., ][]{Bard2008}. Therefore, solving the radiative transfer equation in LTE is not sufficient for our purposes \citep[see, e.g.,][]{Centeno2006, Kuckein2012}, since the emission is at the line core. Thus, we need to solve the radiative transfer problem in NLTE. There are codes available for the solar community, such as the MULTI \citep{Carlsson1986} or the RH \citep{Uitenbroek2001} codes, but these codes are not able to perform inversions. Here we make use of NICOLE \citep{2000ApJ...530..977S,SocasNavarro2015}, which is able to infer the atmospheric parameters through inversions of the Stokes profiles.\ This code has never been used to invert observations of the \sil line so the results presented in this paper concentrate on a single profile alone. A full analysis of the sunspot is presented elsewhere. The first, and more important, ingredient we need to tackle the NLTE problem is a tractable Silicon atom model. The atom model allows us to compute the level populations involved in the \sil line transition. There are several atom models available in the literature, for example, those by \cite{Bard2008} or \cite{Shchukina2012}. Particularly, in the first case, the authors started from a complex and complete model (as was used in the latter work) and developed a simpler model containing the main characteristics of the silicon atom (including the transition of interest). Thus, they constructed a quintessential silicon model that is computationally tractable for NLTE calculations. The use of simpler atomic models is recommended since they simplify the inversion process, which can take tens of hours in a standard computer.  This also speeds up the analysis when one needs to invert a large amount of pixels.

\subsection{NLTE synthesis\label{nlte_synt}}

We have checked whether NICOLE produces meaningful results using the silicon atom model of \cite{Bard2008}. To this end, we computed the intensity spectrum under LTE and NLTE conditions and compared the resulting profiles (see \fig{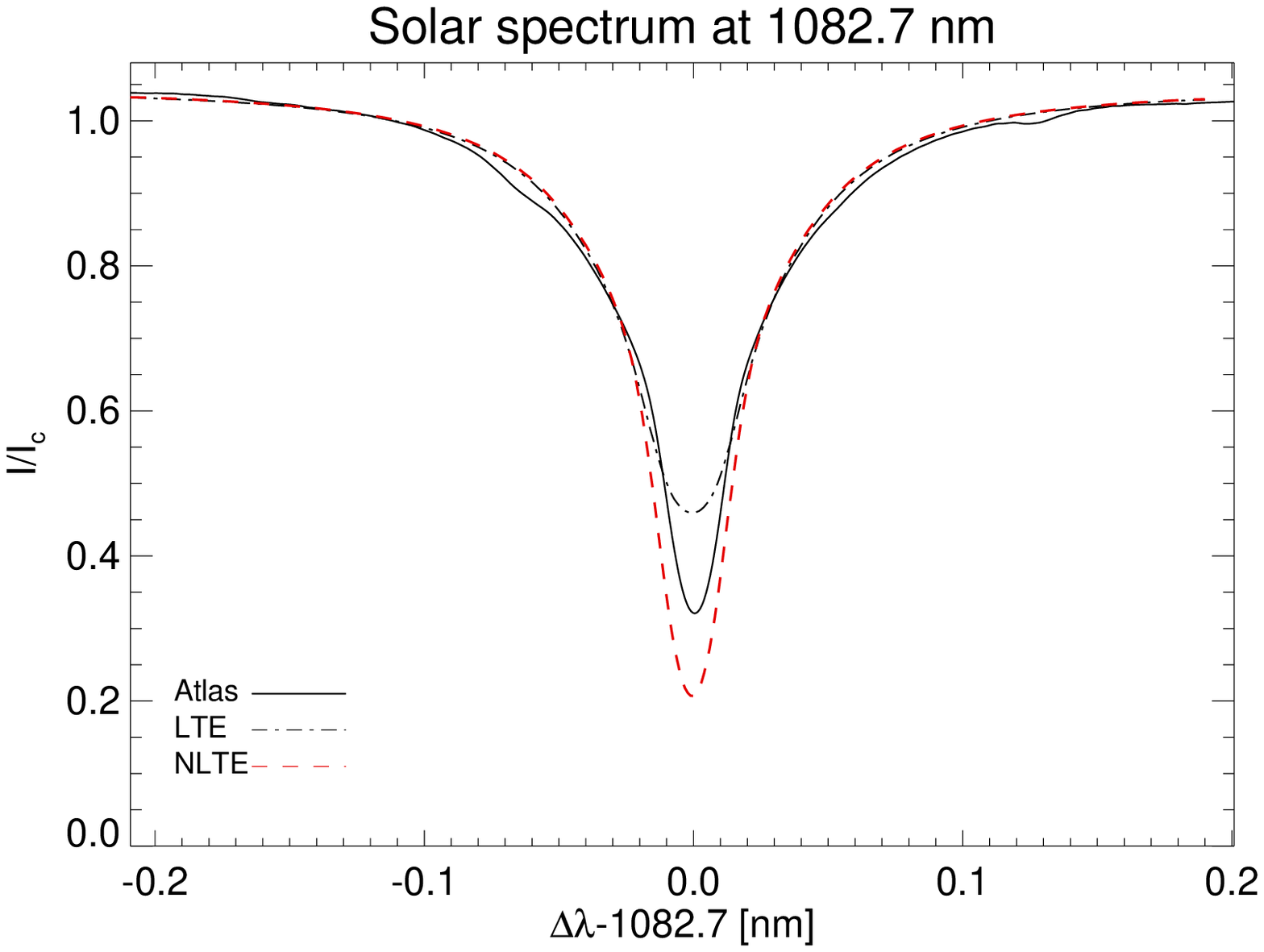}). In the plot, we included the \sil profile taken from the solar atlas of \cite{Delbouille1973}. To set the atomic parameters of the transition, we followed the values listed on Table A.1 of \cite{Borrero2003} including collisional broadening effects \citep{Anstee1995,Barklem1997}. We used the quiet Sun FALC model of \cite{Fontenla1993} to have a fair comparison with the atlas spectrum and also to be able to compare our results with those presented in \cite{Bard2008}. The temperature stratification of this model is plotted in \fig{fig1.eps}. We kept the original microturbulence stratification of the FALC model and added a macroturbulence of 2 km/s to match the typical convective broadening of the lines \citep[][used the same value in their synthesis]{Bard2008}. \Fig{fig6.eps} shows that the \sil line wings are identical in both LTE and NLTE computations, as expected, while there are large differences at the line core wavelengths. In particular, the NLTE profile is much deeper than the LTE profile. A similar result is presented in Figure 3 of \cite{Bard2008}. Regarding the atlas profile, we can see that the NLTE profile is deeper, which could be related to the atomic parameters used here \citep[see the discussion on][]{Bard2008}. We plan to further investigate this particular issue in a future work. These differences were also found on \cite{Bard2008} (see their Figure 18). This simple proof assures us that the atom model has not been perverted during the conversion process.

\colfig{fig6.eps}{Spectral atlas of the Si~{\sc i} line (solid). We include the results of a LTE (dashed) and NLTE (dot-dashed) synthesis using the FALC model atmosphere.\label{Atlas}}

In order to check if the Si~{\sc i} line appears in emission, we synthesized the line profile using the three semi-empirical umbral models presented in \fig{fig1.eps}.
None of the models provided information about the magnetic field vector so we added a constant magnetic field of 2500 G strength and 170 degrees of inclination to mimic the umbral conditions. We chose the latter value to match the observed Stokes V polarity and the small Stokes Q and U signals detected in the umbra. In addition, to simplify this comparison as much as possible we set the microturbulence constant with height and equal to one km~s~$^{-1}$ (except in the FALS model where we kept the original microturbulence stratification). Regarding the macroturbulence, we set it to 2~km~s~$^{-1}$, which is slightly above the  1.41~km~s~$^{-1}$ of macroturbulence that corresponds to the 120~m\AA\/ spectral resolution of GRIS at 1083~nm \citep{2016A&A...596A..59F}. This parameter is kept fixed in the inversion since we expect negligible spatial variations of the macroturbulence in the umbra.

The synthetic Stokes I/I$_\mathrm{c}$ and V/I$_\mathrm{c}$ profiles are shown in \fig{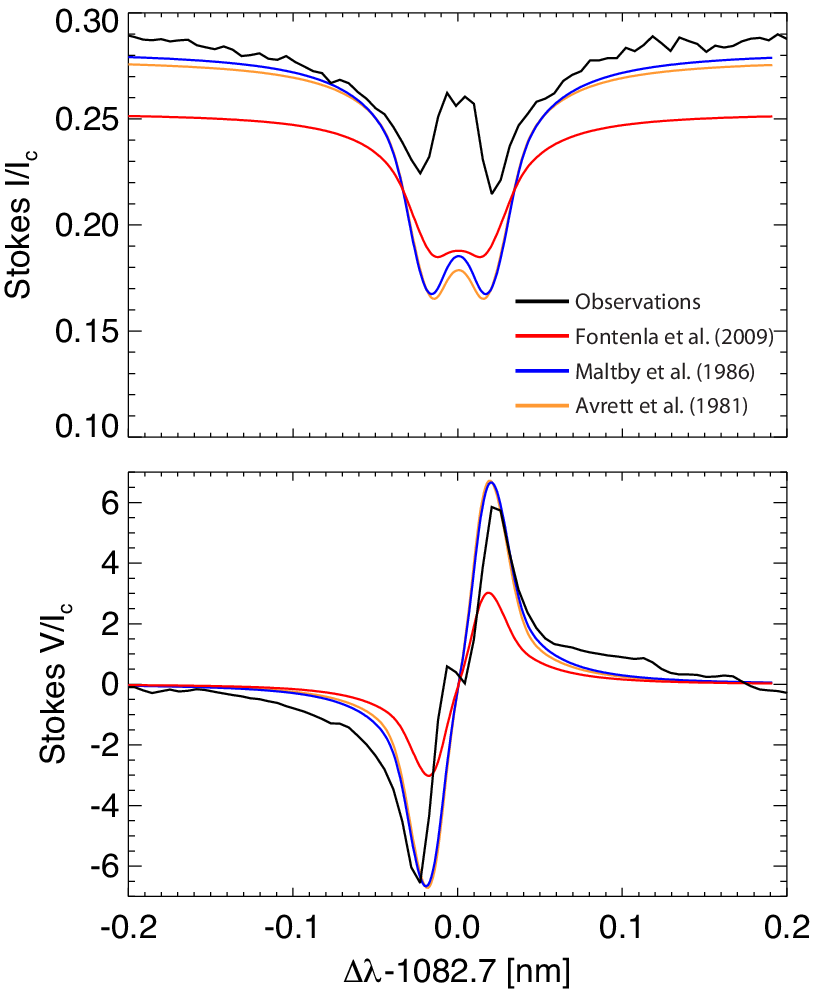}. The Stokes I/I$_\mathrm{c}$ continuum levels of the Maltby and Avrett umbral models match the observed and deconvolved profile with 42\% of scattered light.
 If we consider only 30\% of scattered light, the continuum level is about 0.36 (see second panel in \fig{fig3.eps}), which is far from the synthetized continuum umbral levels. Regarding the Stokes I/I$_\mathrm{c}$ wings, the synthesized profiles show a very similar behavior compared to the observed profile. However, the residual intensity at the line core in the synthesized profiles is much lower than the observed profile, i.e., the line core intensity in the observations is much higher. The two minima in the Stokes I/I$_\mathrm{c}$ synthesized profiles correspond to the well-known Zeeman splitting pattern due to a strong magnetic field. In particular these profiles correspond to the $\sigma$ components, which are located at equidistant wavelength locations form the central wavelength. Their separation depends linearly on the field strength \citep[for instance, see][]{Egidio2004}. If the magnetic field is strictly pointing to the observer there is no central unshifted $\pi$ component. Therefore, the intensity difference between the synthesized profiles and the observed profile at the line core wavelengths ascribed to any other physical mechanisms. 
 
Regarding Stokes V/I$_\mathrm{c}$, none of the models are able to generate a line reversal at the zero-crossing point. We could argue that the Stokes V line reversal appears because Stokes I is in emission at the line core. This would require the temperature to rise at lower atmospheric heights. However, this reasoning is too simplistic. The flanks of the line where the two $\sigma$ components are located are deeper (in intensity units) than the line core and therefore the $\sigma$ components would be ``formed'' higher in the atmosphere than the line core. Thus, if the temperature is rising at lower heights, the $\sigma$ components in both Stokes I and V would be the first to appear in emission and not the line core. In general, the exact physical mechanisms are much more complex, particularly in NLTE conditions. As explained before, we discarded the magneto-optical effects as the source of the Stokes V zero-crossing reversal since the linear polarization signals are very small.

\colfig{fig7.eps}{Synthetic profiles from the semi-empirical models shown in \fig{fig1.eps}. We also included in black the observed and deconvolved profile with 42\% of scattered light shown in the second panel of \fig{fig4.eps} for comparison purposes. The observed profile is normalized to I$_\mathrm{QS}$, which is the average continuum intensity in quiet Sun areas. I$_\mathrm{c}$ stands for the continuum intensity at 500~nm.}

\subsection{NLTE inversion of the \sil line\label{nlte_synt}}

\colfigtwocol{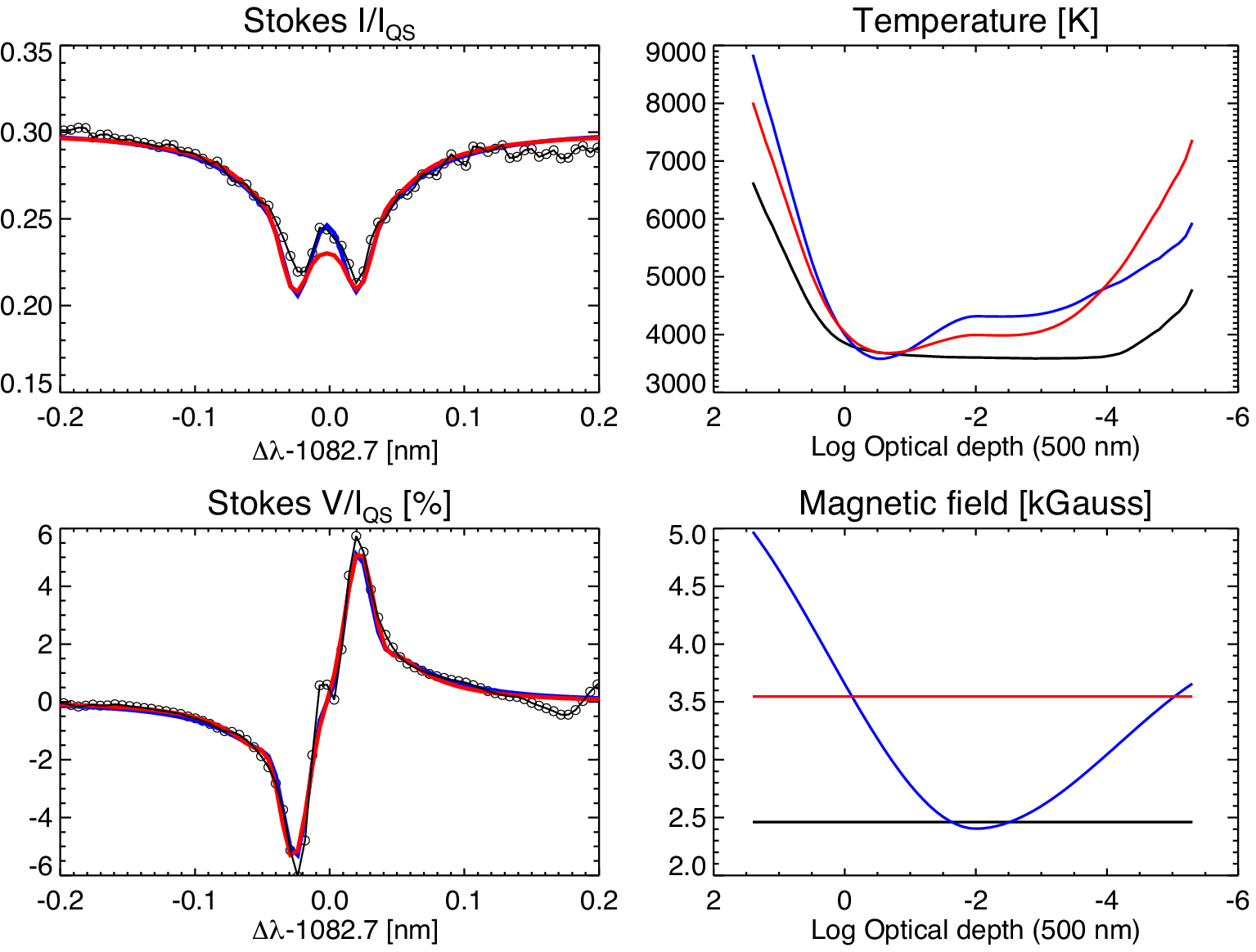}{Left panels: Observed Stokes I and V profiles (open circles connected with black solid line) and their corresponding fits (color) for two of the inversion runs. Right panels: Temperature and magnetic field stratification, corresponding to the initial FALS model, and final inferred model by NICOLE, corresponding to cycles 2 and 3, are shown. The difference between the two cycles is in the number of nodes used for the vector magnetic field, which was a constant for one of these cycles (red) and three nodes for the other (blue).\label{Ajustes}}

We used NICOLE code to invert several selected pixels from the observed map shown in Figure \ref{three}. We started the inversion using the FALS atmosphere as a guess model with a constant magnetic field of 2500~G, 170~degrees of inclination, and 90~degrees of azimuth. In order to invert each profile we performed three iterative cycles. In each one, we increased the number of positions in the atmosphere where the RF are computed (nodes) following the values shown in Table~\ref{nodes}. 

\begin{table}
\vspace{+0.35cm} 
  \begin{adjustbox}{trim=-10 0 10 10,width=0.42\textwidth}
  \begin{tabular}{lccccccc}
        \hline
  Atmospheric parameter      & \vline &  Cycle 1 & Cycle 2 & Cycle 3       \\
        \hline
 Temperature         & \vline  &  1 & 3 & 5     \\      
 LOS Velocity        & \vline  &  0 & 1 & 1    \\
 Microturbulence     & \vline  &  0 & 0 & 1      \\
 Bx                  & \vline  &  0 & 1 & 3     \\ 
 By                  & \vline  &  0 & 1 & 3     \\   
 Bz                  & \vline  &  0 & 1 & 3     \\  
        \hline
  \end{tabular}   
  \end{adjustbox}
\vspace{+0.3cm}
\caption{Nodes configuration used during each cycle of the inversion. }\label{nodes} 
\end{table}

This is the first time NLTE inversions of the Si~{\sc i} have been attempted and we must say that there is additional work that needs to be performed in the future. The most important part is reducing the complexity of the atomic model, if possible. The time required to perform a single inversion cycle ranges from $1\sim8$ hours using an intel i7-6700 CPU whose frequency ranges from $3.7-4.0$~Ghz, i.e., it is included among the fastest frequencies that can be achieved with present CPUs without overclocking them. This long time per cycle means that for computing a full inversion, between 2-3 cycles, almost one day is required. This is more than 100 times slower than the NLTE inversion of the Ca~{\sc ii} 8542 \AA \ line, which can be modeled with a much more simple atom \citep[e.g.,][]{Shine1974}. Thus, it would be desirable to work with simpler atom models, such as that proposed by \cite{2017A&A...603A..98S}, to speed up the initial trial and error analysis before analyzing maps. In any case, NICOLE can work in parallel, i.e., we can speed up the inversion linearly with the number of cores available.
In this section we discuss the results of the best-fitted profile. We would like to emphasize that it was extremely difficult to fit profiles such as those presented in the right-most column of Figure \ref{evolucion}, where the line core intensity is above the continuum level. 

The inversion results corresponding to the second and third cycle are shown in \Fig{fig8.eps}. The NICOLE code fits to Stokes I are rather satisfactory overall. When we look into details, we can see that the line core is not very well fit when we use a model with constant magnetic field (about 3.5~kG and corresponding to cycle 2). The fit is more accurate when we increase the degrees of freedom in the magnetic field (cycle 3). In the latter case, the inferred magnetic field decreases linearly from 3.5~kG at $\log\tau= 0$ to 2.5~kG at $\log\tau= -2$ and then it increases again reaching 3.5~kG at $\log\tau= -5$. 
Regarding the temperature stratification, the inversion, in both cases, tries to introduce a hump around $\log\tau\approx-2$ and an overall increase of the temperature beyond $\log\tau= -2$. Interestingly, none of the inversions are ``apparently'' able to fit the reversal in Stokes V at the zero-crossing point. As we mentioned in \Sect{emission}, the PCA deconvolution method is applied separately to Stokes I and Stokes V. So, there is a possibility of having two Stokes profiles that are not compatible at the central wavelength. For instance, the Stokes V profile, at the zero-crossing wavelength, is usually more affected by noise during the deconvolution process. Indeed, we have seen small variations around the Stokes V zero-crossing point when changing the number of iterations (see Appendix A). Another reason could be that the amount of stray light is too large; the reversal is recovered only when we take 42\% of stray light, as opposed to the emission in Stokes I, which already appears with 30\% levels. However, as we showed before, we do not recover any Stokes V reversal at the zero-crossing point either in the Ca\,\textsc{i}\,1083.34\,nm or Na\,\textsc{i}\,1083.48\,nm and Ca\,\textsc{i}\,1083.9\,nm lines, after applying the same deconvolution procedure and with noisier Stokes V profiles. Finally, the linear polarization profiles were also successfully fit in the inversion process. The inclination of the field is, at all heights, almost vertical (within 170 and 180 degrees).

\section{Summary and discussion \label{discussion}}

In this paper we have shown that, at high spatial resolution and in the absence of scattered light, the \sil line is in emission in sunspot umbrae. To this end, we took observations in the near-infrared spectral range containing the \sil line with the GRIS instrument of the German GREGOR telescope at a spatial resolution of $\approx$~0\farcs3. Using a Mercury transit and the information provided by the adaptative optics system of GREGOR, we modeled the GREGOR PSF during the observations and corrected the data using a novel PCA deconvolution method. The results suggest there is a possibility that the \sil line appears in emission in sunspot umbrae with a line reversal in Stokes V at the zero-crossing point. We have shown how a large amount of stray light coming from wide angles can easily hide the emission peak in the \sil line. Only removing a 30\% of the GREGOR stray light would suffice to start detecting the silicon line in emission. Remarkably, the applied amount of stray light has been determined by a Mercury transit, about 42\%. With this amount of stray light the umbral continua of the recovered profiles match both: the sunspot umbral continuum levels predicted with the current model for a sunspot umbra of about 2.5~Mm radius \citep{refId0,1986ApJ...306..284M} and the continuum level obtained by synthesizing the \sil spectral line with the umbral models of \cite{1981phss.conf..235A} and \cite{1986ApJ...306..284M}. However, the latter models are not able to reproduce the line core intensity after taking into account NLTE effects. Remarkably, neither of the lines nearby the \sil line, i.e., the Ca\,\textsc{i}\,1083.34\,nm and Na\,\textsc{i}\,1083.48\,nm lines, which are both located at the sides of a Ti\,\textsc{i}\,1083.36\,nm atmospheric line, are in emission after the deconvolution. 

We investigated whether the emission is a byproduct of the PCA deconvolution technique; but since we obtained the same results using an alternative deconvolution technique (based in the maximum entropy method), and because of the similarities of the observed GREGOR Stokes profiles with those coming from the German VTT, we think the recovered Stokes profiles are possibly real. Still, the presence of noise in the Stokes V profiles and the number of iterations can slightly alter the results around the zero-crossing point in Stokes V. 

To analyze the \sil spectral line, and because it suffers from NLTE effects around the line core, we adapted the silicon atom model of \cite{Bard2008} to NICOLE and used the latter code to invert the observed Stokes profiles. The NICOLE inversion shows that to get the emission in the \sil line, the temperature stratification should have a hump located at about $\log\tau=-2$ and then should start rising at lower heights when going into the transition region. This result may explain the controversy between the present empirical models and the radio observations at the millimetre range \citep{2016ApJ...816...91I,2015ApJ...804...48I}. The fit is better when we increase the degrees of freedom in the magnetic field strength although the recovered field strength stratification, which first decreases linearly from 3.5~kG at $\log\tau= 0$ to 2.5~kG at $\log\tau= -2$ and then increases again afterward, seems unrealistic; in magnetostatic equilibrium and in a stratified atmosphere (with gas pressure and density decreasing with height) the magnetic field should decrease with height. Remarkably, NICOLE is unable to reproduce the Stokes V line reversal. The reason may be that the reversal is not real either because the noise levels have increased after the deconvolution (see Appendix A) or the amount of stray light is too large; the reversal is recovered only when we take 42\% of stray light, as opposed to the emission in Stokes I, which already appears with 30\% levels. We do not detect the Stokes V line reversal when deconvolving the Ca\,\textsc{i}\,1083.34\,nm and Na\,\textsc{i}\,1083.48\,nm lines, and in this case, the noise in polarization is larger since the signals are much smaller than for \sil.

 In summary, the current results point to a temperature stratification in sunspot umbrae that is much more complex that that provided by current umbral models. The main relevance of these results is that the temperature minimum region may be located at a lower height above sunspot umbrae than in the photosphere. This particular issue has been already discussed by \cite{1994ASIC..433..169S}, where the authors point out that the temperature gradient at the high photosphere should be steeper than in current models.

Our finding provides insights to understand why the sunspot umbra emission in the millimeter spectral range is less than that predicted by current empirical umbral models. However, we need to carry out a more systematic and conscientious investigation of the formation and sensitivities of the \sil line, using numerical simulations and additional observations, to fully understand the formation of this line in strong magnetic field concentrations, such as network, pores, and sunspots. Other effects, such as the influence of the Zeeman splitting in the determination of the level populations or abundance variations, has not been taken into account in the inversion process.  A deep analysis of the sensitivity of the Si I line, including the magnetic field in the statistical equilibrium equations, is a work in progress. The next step will be to analyze the full sunspot, including the Ca\,\textsc{i}\,1083.34\,nm and Na\,\textsc{i}\,1083.48\,nm lines. Thus, the results presented here represent the first step toward the inversion of the \sil line including NLTE effects.

\begin{acknowledgements}
Special thanks to Mats Carlsson (who provided the atom model) and Han Uitenbroek for their support during the NLTE analysis of this work. The 1.5 meter GREGOR solar telescope was built by a German consortium under the leadership of the Kiepenheuer-Institut f\"ur Sonnenphysik in Freiburg with the Leibniz-Institut f\"ur Astrophysik Potsdam, the Institut f\"ur Astrophysik G\"ottingen, and the Max-Planck-Institut f\"ur Sonnensystemforschung in G\"ottingen as partners, and with contributions by the Instituto de Astrof\'isica de Canarias and the Astronomical Institute of the Academy of Sciences of the Czech Republic.
This work was partly supported by the BK21 plus program through the National Research Foundation (NRF) funded by the Ministry of Education of Korea.
This study is supported by the European Commissions FP7 Capacities Programme under the Grant Agreement number 312495.
The GRIS instrument was developed thanks to the support by the Spanish Ministry of Economy and Competitiveness through the project AYA2010-18029 (Solar Magnetism and Astrophysical Spectropolarimetry). 
This work has also been supported by Spanish Ministry of Economy and Competitiveness through projects ESP2014-56169-C6-1-R and ESP-2016-77548-C5-1-R. 
\end{acknowledgements}

\bibliographystyle{aa}
\bibliography{orozco_emission}

\appendix

\section{Variations of deconvolution parameters}

The PCA deconvolution method has two fundamental tuneable parameters: (i) the number of profiles used to construct the database and (ii) the number of iterations used within the Richardson-Lucy deconvolution internal algorithm. The determination of the database size is not trivial. It is usually carried out manually by looking at the profiles. Usually, the user cuts the database at a level where the profiles look like pure Poisson noise. In our case, the level is about $n = 12$. If we decrease the number of profiles in the database ($n <12$) the PCA deconvolution algorithm can miss-reproduce the observed profiles. The number of iterations used for the deconvolution can also alter the final result. Too many iterations can amplify the noise in the final deconvolved profiles. \Fig{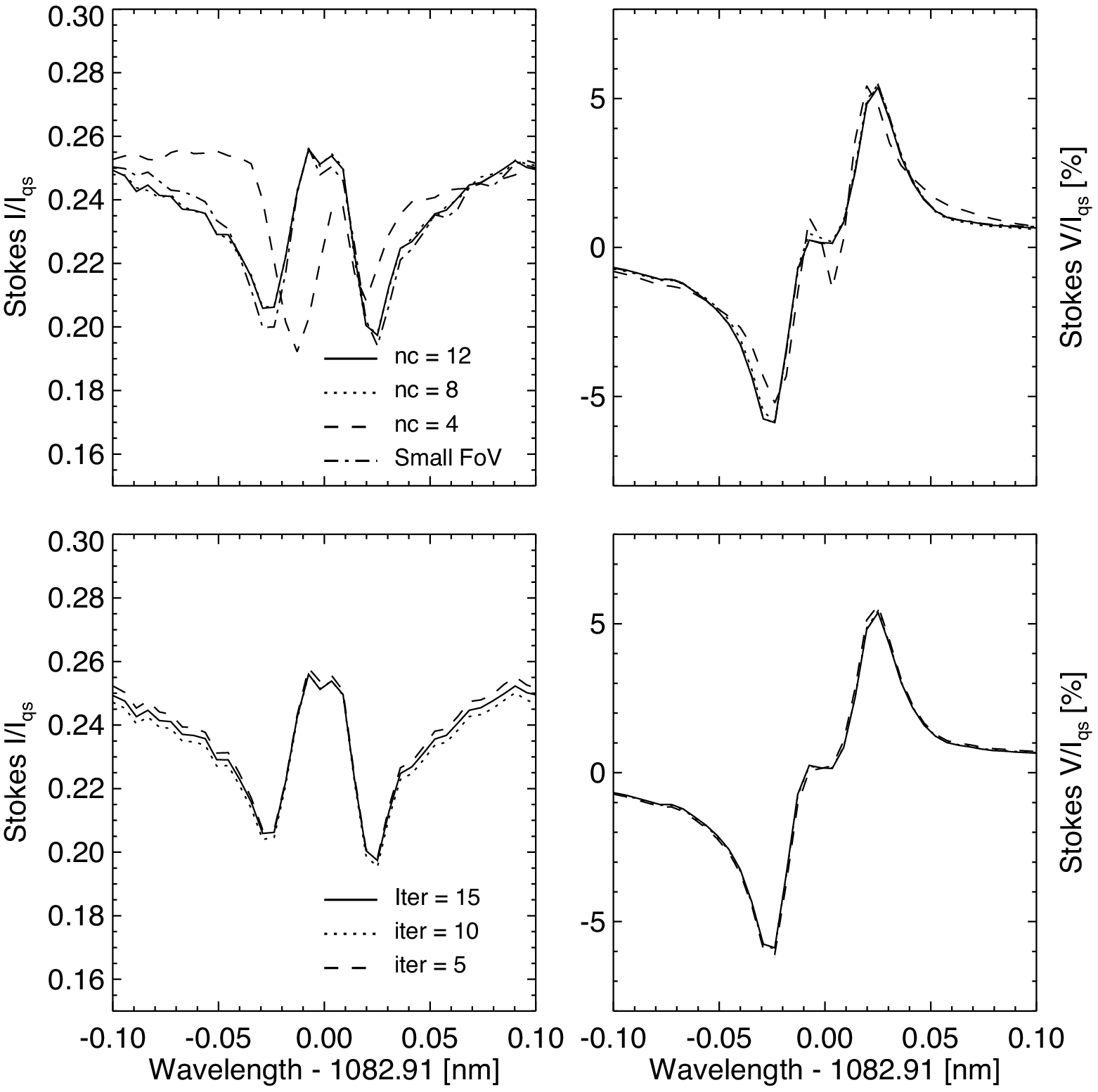} shows how the Stokes I/I$_\mathrm{QS}$ (left panels) and Stokes V/I$_\mathrm{QS}$ (right panels) change for different $n$ (top panels) and number of iterations (bottom panels). It can be seen that Stokes I/I$_\mathrm{QS}$ remains about the same if $n > 8$. For small $n$ values (small database) the recovered Stokes I/I$_\mathrm{QS}$ profile is completely different. In case of Stokes V/I$_\mathrm{QS}$, it is clear that the use of a small database can produce the line reversal at the zero-crossing wavelength point. Regarding the number of iterations, there is no apparent change in Stokes I/I$_\mathrm{QS}$ when the number of iterations varies from 5 to 15, except a slight change in the continuum. In Stokes V/I$_\mathrm{QS}$ there are tiny variations around the zero-crossing points. We also show, in the top panels, the result using a smaller field of view when deconvolving the maps. This is important because the map contains a spot and although the amount of scattered light is constant for every pixel, it is more critical when the image contrast is large (as in a sunspot). 

\colfig{fig1_app.eps}{Variations of a reconstructed umbra Stokes I/I$_\mathrm{QS}$ (left panels) and Stokes V/I$_\mathrm{QS}$ (right panels) profiles when changing the number $n$ of the profiles within the PCA database and the number of iterations.}

\section{Deconvolution of nearby lines}

\Fig{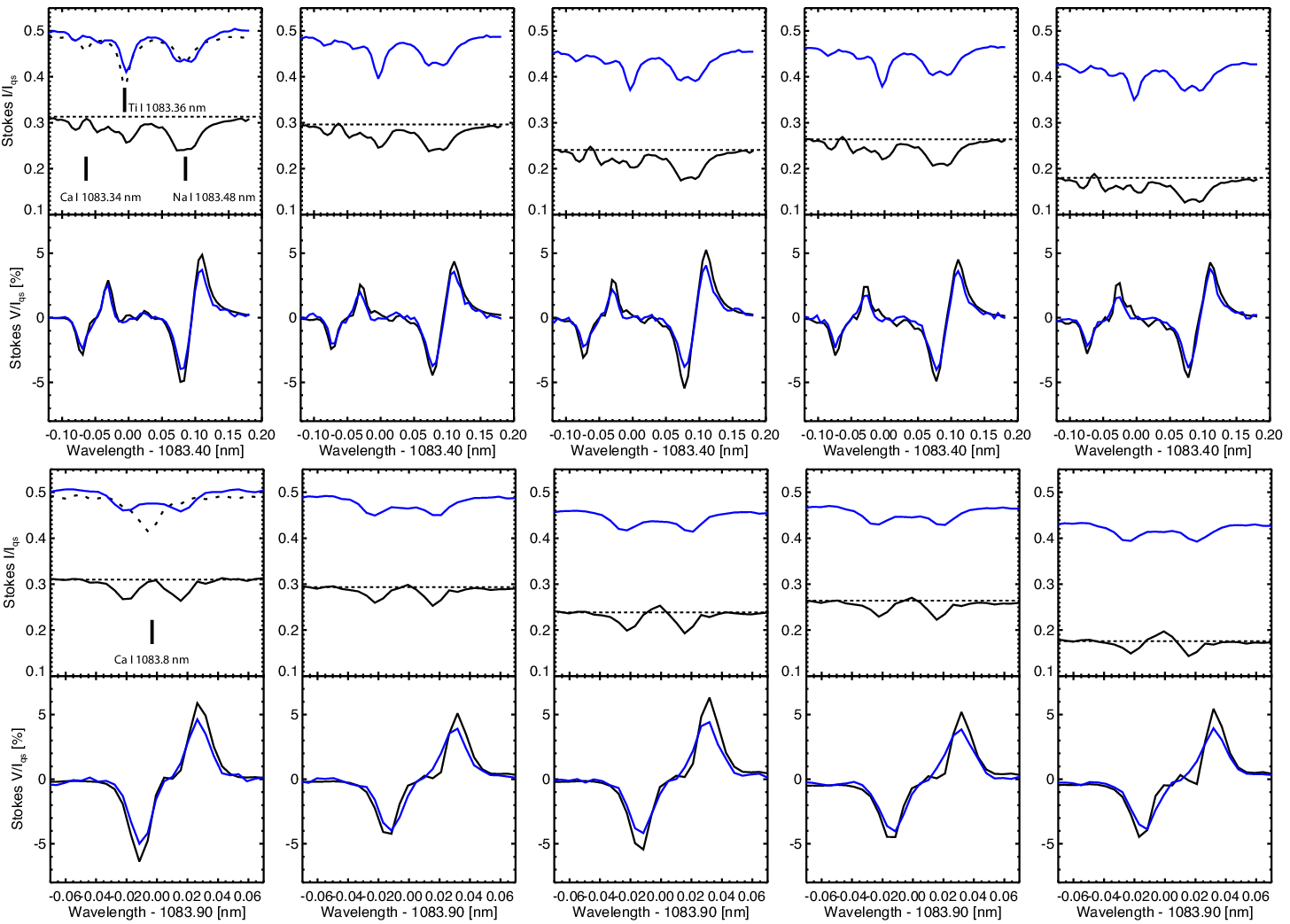} shows the original (blue color) and deconvolved (black color) Stokes I/I$_\mathrm{QS}$ and V/I$_\mathrm{QS}$ profiles corresponding to the Ca\,\textsc{i}\,1083.34\,nm and Na\,\textsc{i}\,1083.48\,nm lines, both located at the sides of a Ti\,\textsc{i}\,1083.36\,nm atmospheric line (top panels) and the Ca\,\textsc{i}\, line at 1083.80\,nm located further in the red (bottom panels). The very first important point to highlight is that none of the Stokes V/I$_\mathrm{QS}$ get a reversal at the zero-crossing wavelengths. In the Ca\,\textsc{i}\, line at 1083.80\,nm we do see some variations in the profile at the umbral core (right most bottom panel) but these are due to noise. In the case of Stokes I/I$_\mathrm{QS}$ it is interesting to see how the core of the Na\,\textsc{i}\,1083.48\,nm line does not reach local continuum values. This is not the case for the Ca\,\textsc{i}\, lines, especially for the Ca\,\textsc{i}\, line at 1083.80\,nm (bottom panels). In this case, the splitting is clearly visible in the line and the residual intensity at the line core of the darker profiles (right most panel) is larger than the continuum. The same happens in the Ca\,\textsc{i}\,1083.34\,nm line, although it is less noticeable. The fact that the Na\,\textsc{i} line  does not behave similarly suggests that the higher continuum intensity values of the line core in the Ca\,\textsc{i}\, lines may be due to the noise. Both Ca\,\textsc{i}\, lines are very shallow. However, in practice we cannot distinguish this effect from an excess of scattered light in the deconvolution. 

\colfigtwocol{fig2_app.eps}{Variation of the original (blue color) and reconstructed (black color) Stokes I/I$_\mathrm{QS}$ and V/I$_\mathrm{QS}$ profiles corresponding to Ca\,\textsc{i}\,1083.34\,nm, Na\,\textsc{i}\,1083.48\,nm, and Ca\,\textsc{i}\, 1083.80\,nm as we move from the inner penumbral boundary (first column) to the center of the umbra (last column). The pixels correspond to those shown in  \fig{fig3.eps}. The dotted profile in the first panel corresponds to a quiet Sun profile (multiplied by x0.5 to fit within the Y-axis range). The horizontal dotted lines are drawn for reference.}

\end{document}